\documentclass{article}
\usepackage[utf8]{inputenc}
\usepackage{natbib}
\usepackage{hyperref}
\usepackage{authblk}

\usepackage{derivative}
\usepackage{amsmath}
\usepackage{mathtools}
\usepackage[makeroom]{cancel} %% to use cross out symbols in equation
\usepackage{subcaption} % Add this package for subfigures
\usepackage{floatrow}
\usepackage[utf8]{inputenc}
\usepackage{graphicx}

\usepackage{float}
\usepackage[
margin=1in
]{geometry}
\setcitestyle{round}
\usepackage{xcolor}
\hypersetup{
    colorlinks,
    linkcolor={red!80!black},
    citecolor={blue!80!black},
    urlcolor={blue!80!black}
}

\renewcommand{\texttt}[1]{\textcolor{magenta}{\ttfamily #1}}

\usepackage{caption}
\usepackage{authblk}

\begin{document}

\title{\vspace{-0.75in} 
% \textbf{Density dependent transition from sessile to collective motion of active dumbbells} \\ 
% or \\
% {\bf Phase-locking driven transition from sessile to collective motion of active dumbbells} \\
% or \\
{\bf Hydrodynamics-driven phase-locking and collective motility of sessile active dumbbells}}
\date{}% It is always \today, today,

\author{Urvi Bora}
\author{Mohd Suhail Rizvi}
\affil{\small Department of Biomedical Engineering, 
Indian Institute of Technology Hyderabad, Kandi Sangareddy, Telangana, 502284, India.}
\maketitle

% \newpage
\begin{abstract}
Collective motion is a phenomenon observed across length scales in nature, from bacterial swarming and tissue migration to the flocking of animals. 
The mechanisms underlying this behavior vary significantly depending on the biological system, ranging from hydrodynamic and chemical interactions in bacteria to mechanical forces in epithelial tissues and social alignment in animal groups. 
While collective motion often arises from the coordinated activity of independently motile agents, this work explores a novel context: the emergence of collective motion in systems of non-motile active agents. 
Inspired by the oscillatory shape dynamics observed in suspended cells such as neutrophils and fibroblasts, we model active dumbbells exhibiting limit-cycle oscillations in shape as a minimal representation of such systems.
Through computational simulations, we demonstrate that hydrodynamic interactions between these dumbbells lead to three key phenomena- a density-dependent transition from sessile to collective motion, hydrodynamics-induced phase separation, and synchronization of oscillatory shape changes. 
We have explored the role of hydrodynamic interactions on these emergent properties of sessile active dumbbells.
These results underscore the critical role of hydrodynamic coupling in enabling and organizing collective behaviors in systems lacking intrinsic motility. 
% Furthermore, our findings bridge theoretical models and experimental observations, providing insights into collective dynamics in biological and synthetic systems. 
This study lays the groundwork for future investigations into the emergent behavior of active matter and its implications for understanding cell motility, tissue dynamics, and the development of bio-inspired materials.

\vspace{1cm}
\noindent {\bf Keywords: }collective motility, active matter, hydrodynamics, phase-locking
\end{abstract}

\newpage

\section{Introduction}

% \texttt{Introduction to collective motion} - 
Collective motion is prevalent in nature at different length scales, ranging from the coordinated migration of cells during embryonic development and collective swimming of prokaryotes to the flocking of birds and animal herds \citep{gueron1996dynamics,toner1998flocks,mehes2014collective,papadopoulou2023dynamics}. 
In different biological contexts, the collective and coordinated nature of motion is essential for organisms' survival and adaptation, enabling processes such as efficient resource acquisition, defense mechanisms, and environmental navigation \citep{berdahl2018collective}.

% \texttt{Mechanisms of collective motion} - 
Since collective motility is observed at diverse lengths and time scales in nature, its underlying mechanisms are also very different and system-dependent. 
For instance, the collective movement of bacterial suspensions is primarily driven by hydrodynamic interactions and chemical signaling, where individual motile bacteria influence each other through fluid flows and gradients of signaling molecules \citep{tian2021collective, be2019statistical}. 
On the other hand, the coordinated movement of epithelial tissues arises from mechanical interactions mediated by cell-cell adhesion, cytoskeletal remodeling, and traction forces generated against the substrate \citep{szabo2010collective, kabla2012collective, camley2017physical}. 
At larger length scales, the flocking observed in larger animals has been attributed to alignment interactions, sensory perception, and social behaviors, where individuals align their movement with neighbors while responding to environmental cues and threats \citep{herbert2011inferring, herbert2016understanding, zheng2024body}.

% \texttt{Individual is an active motile agent} - 
Despite the differences in the underlying mechanisms of the emergence of collective motion, in all of the aforementioned example systems, the individual of the collective has always been observed to be a motile agent \citep{ramaswamy2010mechanics}. 
Therefore, the collective movement is a result of a transition from an uncoordinated movement of individuals to a coordinated movement of the collective. 
For example, a bacterium is capable of swimming on its own with the help of flagellar beating. 
When present in high density the same bacteria can also move collectively resulting in swarming behavior \citep{darnton2010dynamics, be2020phase}.

% \texttt{Context of the current work} - 
In this work, we explore the origin of collective movement in active agents which are inherently incapable of independent motion. 
This investigation is inspired by observed oscillatory shape dynamics in various biological cell types. 
Human neutrophil leukocytes, known for their crawling motility when attached to a substrate, also demonstrate shape oscillations in suspension \citep{ehrengruber1996shape}. 
These cytoskeleton-dependent shape oscillations in these cells in the absence of any underlying substrate have been compared to the formation of lamellipodial and pseudopodial structures that are known to be responsible for crawling cell motility. 
It is not known, however, if these shape oscillations of the suspended neutrophils can result in their swimming motility \citep{farutin2013amoeboid}. 
Similarly, 3T3 fibroblast cells have been observed to exhibit periodic shape oscillations after the loss of adhesion with the substrate \citep{Salbreux_2007}.  
In these fibroblast cells also, the perturbations in actin polymerization or myosin contractility result in the loss of cell shape oscillations.
In addition to the human cells, such periodic shape oscillations have also been observed in \emph{Dictyostelium discoideum} \citep{killich1993locomotion}, \emph{Physarum polycephalum} and \emph{Amoeba proteus} \citep{SATOH198579}. 
These examples show that cytoskeletal activity in the cells suspended in the fluid medium can result in oscillatory shape changes. 

% \texttt{Role of hydrodynamics in collective motility} - 
In the swimming motility of microorganisms, hydrodynamics has been known to be indispensable \citep{lauga2012dance, lauga2016bacterial}. 
For some of the examples of the collective movement also hydrodynamics plays a critical role \citep{ramaswamy2010mechanics}.
As argued by E. M. Purcell in the celebrated `Scallop theorem' the hydrodynamics-driven swimming motility at the microscale requires the swimming agent to perform non-reciprocal changes of the body shapes \citep{Purcell1977}.
Since this constraint requires the microswimmer to have more than a single degree of freedom, one of the simplest models of the microswimmer have been constructed with three identical beads connected with two linear springs in a colinear fashion \cite{C4SM02611J}. 
In contrast, two beads connected with a single spring, also known as a `dumbbell', are incapable of swimming at the microscale, thanks to a single degree of freedom in shape changes. 
The bead-spring assemblies have been used extensively to model polymers \citep{setaro2019dumbbell}, flagellated microswimmers \citep{rizvi2018size}, and cells in extracellular matrix \citep{tsingos2023hybrid}. 

% \texttt{Activity of the cell} - 
Different modeling approaches for bead-spring and other types of microswimmers consider oscillatory shape changes with a fixed time period \citep{C4SM02611J, rizvi2018size, PhysRevE.97.023102}. 
Several experimental observations, however, have demonstrated that depending on the microenvironment and swimmer anatomy the time period might not remain fixed. 
One prominent example of this is the dependence of the beating frequency of the swimmer flagella on fluid viscosity and flagellar length \citep{khona2013anomalies}. 
Cytoskeletal dynamics-driven changes in cell shape and size have also been shown to be modulated by the microenvironment with cell deformation to demonstrate limit-cycle oscillations \citep{PhysRevLett.113.148102}. 
The limit-cycle oscillations of cell shape changes can account for the influence of the surrounding fluid medium, leading to oscillations with time periods dependent on microswimmer properties and microenvironment. 
Similarly, the limit-cycle oscillations with hydrodynamic interactions have also been shown to be the underlying mechanism of synchronization of flagellar beating in microorganisms \citep{Flagellar2014Brumley, Generic2011Uchida, Man2020Multisynchrony}. 

% \texttt{In this work} - 
In this work, we demonstrate the emergence of the collective mode of motility of active dumbbells which are, thanks to the scallop theorem, sessile in isolation. 
% We model oscillatory changes in the dumbbell shape as a cytoskeletal dynamics-dependent limit-cycle oscillator. 
We show that the effect of hydrodynamic coupling between two dumbbells manifests itself in two ways - first, in their collective motion and, second, in synchronization of their shape oscillations. 
We also observe hydrodynamics-induced phase separation of the active dumbbells.

% \begin{enumerate}
%     \item Collective motion \citep{bardfalvy2024collective,sokolov2012physical}
% \end{enumerate}
% Synchronization
% \begin{enumerate}
%     \item Hydrodynamic synchronization at low Reynolds number \citep{golestanian2011hydrodynamic}
%     \item Two colloidal particles under the oscillatory optical trap. Hydrodynamically connected. \citep{kotar2010hydrodynamic}
%     \item Synchronization in Janus particles leads to their self-assembly \citep{yan2012linking}
%     \item Flagellar synchronization through direct hydrodynamic interactions \citep{Flagellar2014Brumley,Generic2011Uchida,Man2020Multisynchrony}
%     \item Synchronization and metachronal waves \citep{Hydrodynamic2012Brumley,brumley2015metachronal}
% \end{enumerate}
% Phase separation
% \begin{enumerate}
%     \item 
% \end{enumerate}
\section{Mathematical Model}

\subsection{Basic assumptions}
We consider biological cells with elongated morphology and model them as dumbbells, each made up of two identical beads connected by a spring (details below).
Here we consider only the deterministic dynamics of the dumbbells and ignore any stochastic effect such as diffusion of the dumbbells. 
% Even though at the sub-cellular level the stochastic effects play crucial roles in normal cell physiology.
This simplification is justifiable in the context of biological cells for which the thermal effects are small as compared to the force generated by the actomyosin cytoskeleton. 
For simplicity, we consider a one-dimensional arrangement of the dumbbells suspended in a three-dimensional fluid medium.
In principle, the assumptions of elongated cell morphology and the one-dimensional arrangement of the dumbbells can be relaxed to three dimensions but we are not considering it here. 
Given the microscopic nature of the biological cells, the Reynolds number associated with the fluid flow generated by the dumbbell activity is assumed to be negligible. 
The flow generated by one dumbbell can, however, influence the dynamics of another dumbbell via hydrodynamic interactions. 
Further, at high concentrations of the dumbbells, the beads belonging to two adjacent dumbbells can come very close to each other. 
To prevent dumbbells from crossing each other, we have also assumed soft repulsion between the beads. 
\subsection{Bead-spring dumbbell as a model of a biological cell}
We model an elongated cell as a bead-spring dumbbell composed of two identical beads (each with radius $a$) connected by a spring. 
This representation of a biological cell is motivated by the model presented in \citep{PhysRevLett.113.148102} for active contractile element. 
The spring connecting the two beads is assumed to be nonlinear with cubic nonlinearity with the spring response force given by 
\begin{equation}
    f_s (l_i) = k_1 \left(l_i-l_0 \right)  + k_3 \left(l_i-l_0 \right)^3 \label{eq:fsli}
\end{equation}
where $l_i$ and $l_0$ are the lengths of the spring in deformed and undeformed configuration, respectively, and $k_1$ and $k_3$ are the elastic constants. 
The activity of the dumbbell is assumed to arise from the force producing myosin motor proteins. 
Myosin proteins bind and unbind to the actin network of the cell such that the rate of change in the concentration of bound myosin for each dumbbell is given by 
\begin{equation}
    \frac{dc_i}{dt} = -\frac{1}{\tau} \left( c_i - c_0 \right) - \frac{c_i}{l_i} \frac{dl_i}{dt} \label{eq:dcidt}
\end{equation}
where the first term on the right-hand side corresponds to the myosin binding and unbinding from the actin network (with a characteristic time $\tau$), and the second term is due to the conservation of myosin. 
The bound myosin generates a contractile force $f_m(c_i)$ in each dumbbell. The length of the dumbbell changes in response to the contractile and restoring forces due to myosin and spring, respectively.
We consider the dumbbell to be suspended in a Newtonian fluid of viscosity $\eta$ and its beads follow overdamped dynamics.
This gives (in the absence of hydrodynamic interactions)
\begin{equation}
    \frac{dl_i}{dt} = \mu \left( f_m(c_i) + f_s(l_i) \right) \label{eq:dlidt}
\end{equation}
where $\mu = \left[ 6 \pi \eta a\right]^{-1}$ is the mobility of the bead. 
As shown in \citep{PhysRevLett.113.148102} this system demonstrates Hopf bifurcation and the resulting limit cycle of the oscillations has time period $T=\sqrt{\tau/\mu k_1}$ at the bifurcation which does not depend on myosin levels. 
Far away from the bifurcation, however, the oscillation time period depends on other system parameters as well. 
Here we have looked at the collective dynamics of these active dumbbells in the presence of hydrodynamic interactions. 
\begin{figure} [H]
    \centering
    \includegraphics[width=0.95\textwidth]{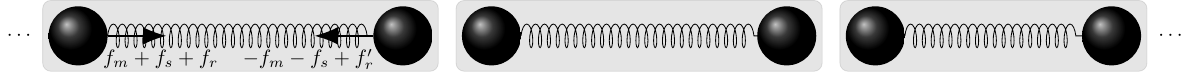}
    \caption{Schematic showing the one-dimensional arrangement of the dumbbells, a representation of an active sessile agent. 
    The arrows in the first dumbbell denote the internal ($f_m(c_i)$ due to myosin contractility and $f_s(l_i)$ passive spring force) and external ($f_r$ repulsion between two beads) forces acting on the two beads. }
    \label{fig:schematic}
\end{figure}
\subsection{Active dumbbell suspension}
We consider a one-dimensional axial arrangement of the active dumbbells suspended in a fluid. 
For this system, we define suspension density
\begin{equation}
    \Phi = \frac{\left( l + 2a \right) N}{L}
\end{equation}
where $N$ is the number of dumbbells in the system and $L$ is the system size. 
The interaction of the dumbbells with the fluid is only through the beads and the spring does not interact with the fluid medium.
In this dumbbell suspension, each bead experiences forces of three nature- (i) intrinsic passive forces due to the nonlinear spring, (ii) intrinsic active force due to the myosin activity (both described in the previous section), (iii) external repulsion force 
\begin{equation}
    f_r(r) =   \begin{cases} 
                    \cfrac{\epsilon}{r^n} & \text{if } r \leq 2a \\
                    0 & \text{otherwise} 
                \end{cases} 
\end{equation}
from other beads, where $r$ is the center-to-center distance between two beads and $\epsilon$ is a phenomenological constant.  

The movement of the beads can generate flow in the fluid. 
As mentioned earlier, due to the microscopic nature of such systems, the inertia in the fluid flow is negligible resulting in Reynolds number $R_e \sim 0$. 
For this condition, the fluid flow can be described by the Stokes equations 
\begin{equation}
    \nabla \cdot \mathbf{u} = 0; \eta \nabla^2 \mathbf{u} - \nabla p = 0
\end{equation}
and the equation of motion of a bead is 
\begin{equation}    \frac{d r_{\alpha}}{dt} = \mu f_{\alpha} + \sum \limits_{\beta \ne \alpha} G_{\alpha \beta} f_{\beta}  \label{eq:dradt}
\end{equation}
where indices $\alpha$ and $\beta$ indicate bead identity, $f_{\alpha}$ is the total force acting on $\alpha$-th bead,  and
\begin{equation}
    \mathbf{G}_{\alpha \beta} = \frac{1}{8 \pi \eta} \left( \frac{\mathbf{I}}{r_{\alpha \beta}} + \frac{\mathbf{r}_{\alpha \beta} \otimes \mathbf{r}_{\alpha \beta}}{r_{\alpha \beta}^3}\right)
\end{equation}
is the Oseen tensor between two beads in three dimensions. 
% For the one-dimensional arrangement of the dumbbells studied in this paper, it simplifies to 
% \begin{equation}
    % G_{\alpha \beta} = \frac{1}{ 4 \pi \eta r_{\alpha \beta}}.
% \end{equation}
We solve equations of motion for each bead (equation \eqref{eq:dradt}) and track the centers of mass of individual dumbbells and the collective for analysis. 
\subsection{Non-dimensionalization}
The system has two characteristic lengths- equilibrium spring length $l_0$, and bead-radius $a$. 
We non-dimensionalize the lengths in this work by the average equilibrium length of the spring $l_0$. 
Given the hydrodynamic and cytoskeletal dynamics in the system there are several different characteristic times present in the system. 
For instance, we have dumbbell relaxation time $\cfrac{k_1 a^2}{\eta}$, myosin turnover time $\tau$, myosin driven contractility time $\cfrac{t_1 a^2 c_0 }{\eta}$, to list a few. 
Since this model of dumbbell shows limit cycle oscillations, we have used the time period of a single dumbbell as the characteristic time for the system. 
Therefore, the single dumbbell translational velocity and the collective velocity of the system are measured in the units of $l_0/T_0$, that is dumbbell length in the duration of one oscillation (see Fig. \ref{fig:singleDumbbellDynamics}C).  
% \begin{enumerate}
    % \item Lengths $l$, $a$, $L$
    % \item concentration $c_0$
    % \item time $t_1 a^2 c_0 /\eta$
% \end{enumerate}

\section{Results}
\subsection{Single dumbbell: oscillations and motility}\label{sec:singleDumbbell}
\begin{figure} [H]
    \centering
    \includegraphics[width=0.7\textwidth]{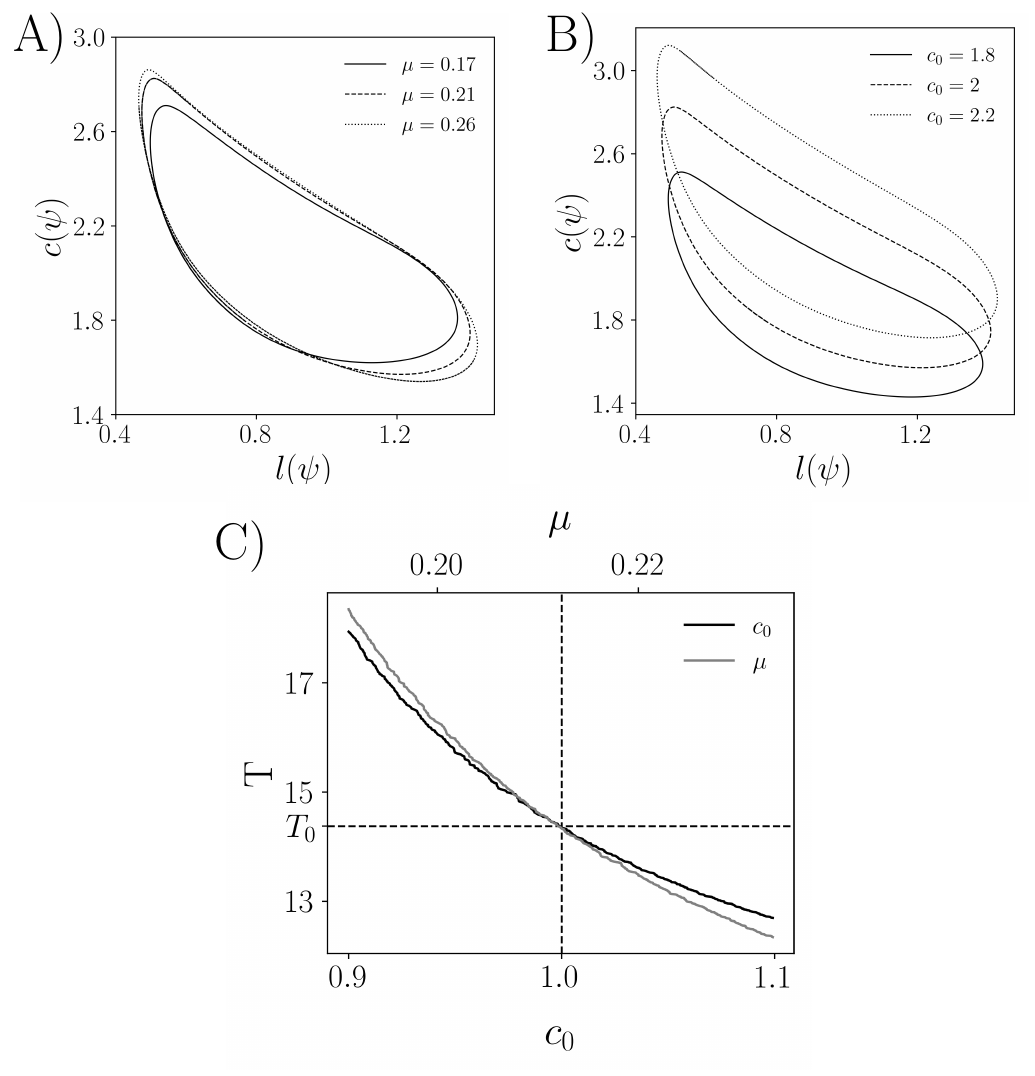}
    \caption{{\bf Single dumbbell dynamics.} Limit cycle oscillations of the dumbells for different values of (A) bead mobilities $\mu$, and (B) reference myosin levels $c_0$. 
    (C) Dependence of time periods of limit cycle oscillations on $\mu$ and $c_0$. 
    $T_0$, the time period for $c_0=1.0$ and $\mu=$ is taken as the characteristic time for the analysis shown in this work. }
    \label{fig:singleDumbbellDynamics}
\end{figure}

For a single active dumbbell in fluid, the myosin activity-induced contractile forces can result in oscillatory dynamics where the length of the dumbbell and the bound myosin concentration oscillate over time. 
In this work, we have considered dumbbells with different reference myosin levels $c_0$ and bead mobilities $\mu$ to study their effect on the collective dynamics of the active dumbbells. 
Figs. \ref{fig:singleDumbbellDynamics}A and \ref{fig:singleDumbbellDynamics}B show state-space trajectories of a single dumbbell for different values of bead mobility $\mu$ and reference myosin concentrations $c_0$. 
Due to the oscillatory nature of the dumbbell (as shown in Fig. \ref{fig:singleDumbbellDynamics}A-B), the configuration of an active dumbbell can be described in terms of a phase variable $\psi$ such that the bound myosin concentration and dumbbell length are functions of this phase variable, that is $c(\psi)$ and $l(\psi)$. 
% The state-space trajectories show the oscillatory dynamics of the dumbbell with oscillation time-periods being dependent on $\mu$ and $c_0$. 
Fig. \ref{fig:singleDumbbellDynamics}C shows the dependence of the time period of limit cycle oscillations of an active dumbbell on $\mu$ and $c_0$. 
It demonstrates faster oscillations with increasing bead mobility as well as high reference myosin levels. 

% Given the oscillatory nature of the dumbbell (as shown in Fig. \ref{fig:singleDumbbellDynamics}A-B), the configuration of an active dumbbell can be described in terms of a phase variable $\psi$ such that the bound myosin concentration and dumbbell length are functions of this phase variable, that is $c(\psi)$ and $l(\psi)$.
With the help of phase variable $\psi$, as an approximate description of the steady state oscillatory nature of the dumbbell we can write 
\begin{align}
    l(\psi) &= \bar{l} + \tilde{l} \sin(\psi) \\ 
    c(\psi) &= \bar{c} + \tilde{c} \sin(\psi + \delta)
\end{align}
with $\psi = \omega t$. 
We will use this description to study the effect of hydrodynamic interactions between two active dumbbells analytically.

This oscillatory dynamics, however, does not generate any translational movement of the dumbbell at a low Reynolds number. 
This is not surprising since the oscillatory dynamics of the dumbbell results in reciprocal changes in its configuration over time reversal \citep{Purcell1977}. 
A single dumbbell can demonstrate translational motion if it is in the vicinity of another dumbbell with which it can interact hydrodynamically.

\subsection{Hydrodynamic interactions and collective motion}

\subsubsection{Pair of active dumbbells: an analytical argument}\label{sec:pair_of_active}
To study the effect of hydrodynamic interaction on the translational motion of dumbbells we first looked at a pair of active dumbbells arranged in an aligned configuration. 
As described earlier, the oscillatory dynamics of the two dumbbells can be described by 
\begin{align}
    l_1(t) &= \bar{l}_1 + \tilde{l}_1 \sin(\omega_1 t) \\
    c_1(t) &= \bar{c}_1 + \tilde{c}_1 \sin(\omega_1 t + \delta_1) \\ 
    l_2(t) &= \bar{l}_2 + \tilde{l}_2 \sin(\omega_2 t + \theta) \\
    c_2(t) &= \bar{c}_2 + \tilde{c}_2 \sin(\omega_2 t + \theta + \delta_2)
\end{align}
where $\theta$ is the phase difference between the two dumbbells. 
Please note that to write the above expressions we have ignored the effect of inter-dumbbell interaction on myosin binding-unbinding. 
In the next section, we will relax this constraint and analyze the system numerically. 

If the two dumbells are at a center-to-center separation $d$ (where $d \gg l$) from each other, the system can be analyzed analytically for $\bar{l}_i \gg \tilde{l}_i$. 
We obtain for $\omega_1=\omega_2=\omega$ (leading order terms in $\tilde{l}_i$)
\begin{align}
    v_1 + v_2 &\sim \frac{k\left( \tilde{l_2}^2 \cos \theta - \tilde{l_1}^2\right)}{4 \eta l \omega r^2}\label{eq:pairVsum}\\ 
    v_1 - v_2 &\sim \frac{k\left( \tilde{l_2}^2 \cos \theta + \tilde{l_1}^2\right)}{4 \eta l \omega r^2}
\end{align}
where $v_i$'s are the translational velocities of each dumbbell averaged over one oscillatory cycle. 
More general expression for $\omega_1 \ne \omega_2$ can also be obtained but it too long to report here. 
% Please note that the expressions above are only approximate as they contain only the leading order terms of the velocity expressions. 
This shows that in the presence of hydrodynamic interactions between the active dumbbells, they can collectively move together as long as they do not oscillate synchronously. 
Further, the separation between the two dumbbells also does not remain fixed over time as $v_1 - v_2 \ne 0$ for all values of $\theta$. 
This implies that depending on the oscillatory phases of the two dumbbells 
% and the difference in the sizes of the beads in the two dumbbells 
they can come closer to (or move away from) each other. 
This demonstrates a phase-dependent effective attraction or repulsion between two hydrodynamically coupled active dumbbells. 
This analytical argument suggests that for a large number of active dumbbells with different oscillatory phases, bead sizes, and time periods, we expect to see collective motility and formation of dumbbell clusters due to hydrodynamics-driven effective attraction and repulsion. 
\subsubsection{Collective motility: numerical simulations}
\begin{figure} [H]
    \centering
    \includegraphics[width=0.9\textwidth]{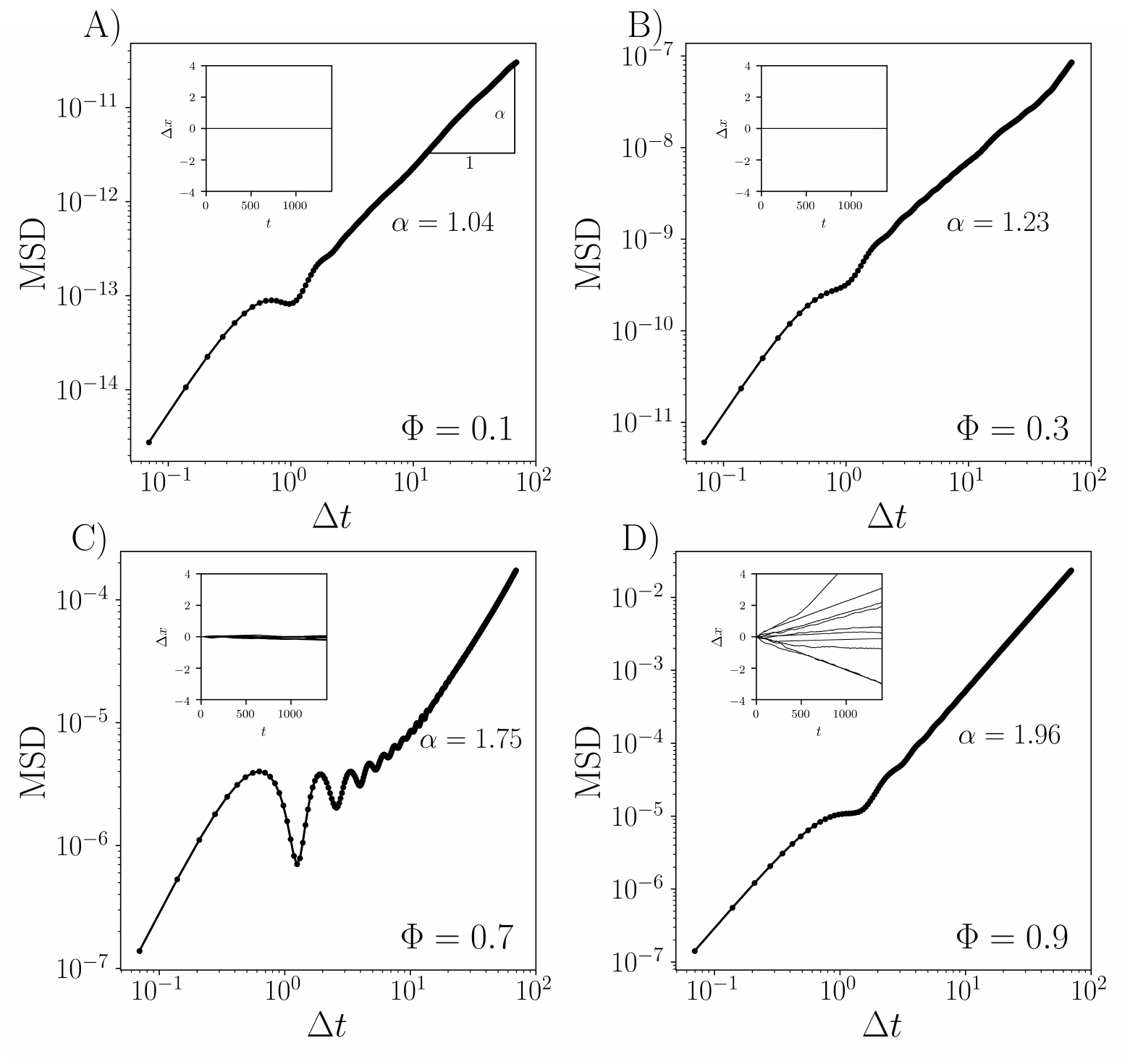}
    \caption{{\bf Collective motion of dumbbells. }Mean square displacement (MSD) of the center of mass of active dumbbell suspension for different suspension density values.
    The exponent $\alpha$ increases with increasing $\Phi$. 
    Insets show the positions of the system's center of mass as a function of time for different realizations.}
    \label{fig:second-figure-msd}
\end{figure}

Next, we numerically solved the equations of active dumbbell dynamics (equations \eqref{eq:dcidt}-\eqref{eq:dradt}) for a system containing $N=64$ dumbbells at different suspension densities. 
The dumbbells were placed randomly in a one-dimensional arrangement (as shown in Fig. \ref{fig:schematic}) with periodic boundary conditions. 
We varied suspension density from $\Phi=0.1$ to $\Phi=0.9$ and calculated the mean-square-displacement (MSD) of the center of mass of all dumbbells in the suspension. 
Four representative MSD plots are shown in figure \ref{fig:second-figure-msd}. 
It can be seen that at low suspension density ($\Phi=0.1$) the center of mass of the system does not show any overall translational movement (inset in Fig. \ref{fig:second-figure-msd}A). 
At low $\Phi$ even though individual dumbbells may be translating the averaged translational velocity vanishes (Fig. \ref{fig:Histogram_velocity_sum_diff}A). 
A calculation of the MSD for this case shows cases with sub-diffusive and super-diffusive motions of the center of mass of the suspension. 
Still, an average over several simulations shows $\text{MSD} \sim \Delta t^{\alpha}$ with $\alpha \approx 1$, a signature of diffusive behavior. 
At high density ($\Phi=0.9$), on the other hand, the center of mass demonstrates a persistent translational movement with ballistic or super-diffusive signatures ($\alpha \approx 2$) (Fig. \ref{fig:second-figure-msd}D). 

\begin{figure} [H]
    \centering
    \includegraphics[width=0.88\textwidth]{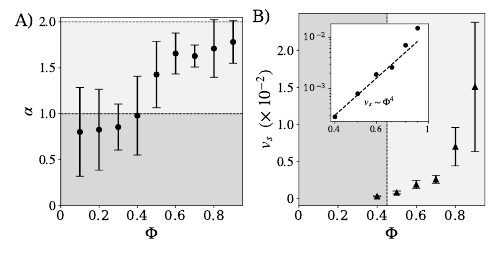}
    \caption{
    (A) Exponent $\alpha$ (as defined in Fig. \ref{fig:second-figure-msd}) in the mean square displacement plot for different values of $\Phi$.
    (B) Velocity of the system's center of mass for different suspension density values from repeated numerical simulations for each $\Phi$. 
    The error bars denote standard deviations.
    The inset shows the mean velocity values on a log-log plot showing a scaling relation $v_s \sim \Phi^4$.  
     }
    \label{fig:slopeandvelocity}
\end{figure}

From the numerical simulations for different $\Phi$, we quantified exponent $\alpha$ and average translational velocity $v_s$ of the center of mass for the cases showing ballistic motion. 
Fig. \ref{fig:slopeandvelocity} shows the dependence of $\alpha$ and $v_s$ on suspension density. 
We observe a transition from diffusive ($\alpha \approx 1$) to ballistic ($\alpha \approx 2$) motion of the center of mass of the system as the suspension density is increased (Fig. \ref{fig:slopeandvelocity}B). 
With an increase in the suspension density, we see the emergence of collective motion at $\Phi \approx 0.4$ for the center of mass of the dumbbells (Fig. \ref{fig:slopeandvelocity}A). 
A log-log plot of the suspension center of mass velocity with density demonstrates a power law relation $v_s \sim \Phi^4$ (inset in Fig. \ref{fig:slopeandvelocity}B). 
From the analytical argument presented in section-\ref{sec:pair_of_active} the collective translational velocity of a pair of dumbbells is expected to scale as $v_s \sim \Phi^2$ for small values of $\Phi$.
For a large number of dumbbells, on the other hand, the average collective velocity $v_s \sim \Phi^2 \langle \cos \theta_{i,i+1} - 1 \rangle_{i} \sim 0$ where the $\langle \cdot \rangle_i$ is the average value over all consecutive pairs of dumbbells. 
This explains the absence of collective motion at small $\Phi$. 

% This nature of the dependence of collective velocity on suspension density (for dilute case) can be explained by the following. 
% For an equispaced arrangement of dumbbells, the separation between two dumbbells and the suspension density are related as $\Phi \sim 1/d$. 
% Since a pair of two active dumbbells can collectively move with a velocity proportional to $1/d^4$ (equation \eqref{eq:pairVsum}), it results in $v_s \sim \Phi^4$. 
At high dumbbell concentrations, the observed relation of $v_s \sim \Phi^4$ hints towards the presence of some additional phenomenon, along with the hydrodynamic interactions among the dumbbells. 
In the following, we will first analyze the relative dumbbell motions before exploring the possible mechanism for the emergent collective motion at high densities.  

% \textcolor{red}{The trajectories of individual dumbbells in a suspension demonstrate ... TO BE COMPLETED (Fig. \ref{fig:Histogram_velocity_sum_diff}A). }

% At high-density values ($\Phi=0.8$, $0.9$) in a few realizations, the active dumbbell suspensions demonstrate collective motion at high velocities (Fig. \ref{fig:slopeandvelocity}A). 

\subsubsection{Phase separation}
\begin{figure} [H]
    \centering
    \includegraphics[width=0.7\textwidth]{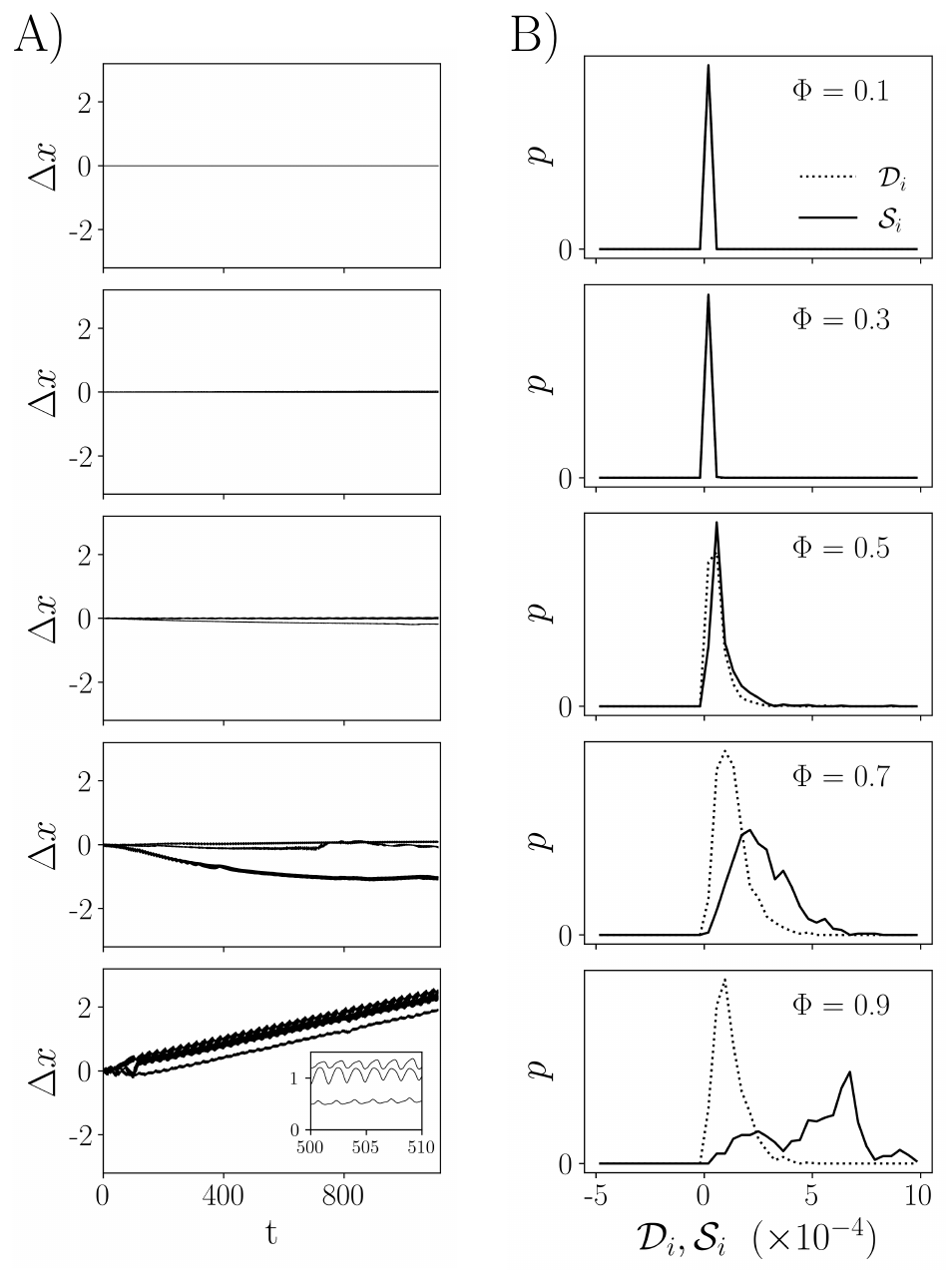}
    \caption{{\bf Collective swimming velocity.} (A) Displacement of representative dumbbells for different values of $\Phi$. 
    (B) Probability distribution of collective $\mathcal{S}_i$ and relative $\mathcal{D}_i$ velocities of adjacent dumbbells.  
    Inset in the bottom panel in (A) shows the enlarged version of the same plot. }
    \label{fig:Histogram_velocity_sum_diff}
\end{figure}

\begin{figure} [H]
    \centering
    \includegraphics[width=0.9\textwidth]{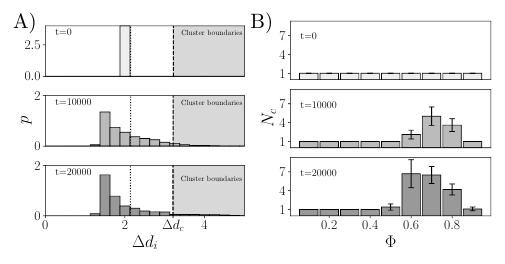}
    \caption{{\bf Hydrodynamic phase separation.} 
    (A) Distribution of relative positions of adjacent dumbbells $\Delta d_i$ for one realization of suspension at $\Phi=0.7$ at different time points.
    Dotted lines mark $\langle \Delta d_i \rangle$ the average value of $\Delta d_i$ for the particular sample. 
    We have used relative separation between adjacent dumbbells exceeding $\Delta d_c = 1.5 \times \langle \Delta d_i \rangle$ as the boundary between two clusters of dumbbells (marked with the dashed lines). 
    (B) Number of clusters for different $\Phi$ at three time points for a 64 dumbbell system. }
    \label{fig:numberDensityEvolution}
\end{figure}

As mentioned earlier (section \ref{sec:pair_of_active}), depending on the phase difference between two dumbbells, hydrodynamic interaction also results in effective attraction or repulsion between them. 
This effective attraction or repulsion can lead to the formation of high and low-density phases in the suspension.
To identify this nature of the active dumbbell suspension, we calculated the absolute difference and sum of the velocities of adjacent dumbbells as 
\begin{align}
    \mathcal{D}_i &= \left| v_{i} - v_{i+1}\right| \\
    \mathcal{S}_i &= \left| v_{i} + v_{i+1}\right|
\end{align}
where $v_i$ is the velocity of the center of mass of the $i$th dumbbell. 
A non-zero value of $\mathcal{D}_i$ indicates two dumbbells moving in opposite directions resulting in formation of high and low-density regions in the suspension near the location of $i$th dumbbell.
On the other hand, a non-zero value of $\mathcal{S}_i$ indicates the collective motion of the two neighboring dumbbells. 
Fig. \ref{fig:Histogram_velocity_sum_diff}B shows the probability distributions of these two quantities. 
It is apparent that at low densities the rates of collective movement and phase separation are negligible. 
However, at high dumbbell densities, we observe an increase in the velocity of collective movement as well as in the rate of phase separation. 
To explore the nature of phase separation and its dependence on the suspension density we calculated relative separation $\Delta d_i =|x_i^c - x_{i+1}^c|$ between the centers of two adjacent dumbbells at different time points. 
Fig. \ref{fig:numberDensityEvolution}A shows the probability distribution of the relative separation between adjacent dumbbells. 
Starting from the equispaced distribution of the dumbbells (at $t=0$) with increasing time, the separation between adjacent dumbbells becomes inhomogeneous leading to cluster formation. 

We defined $\Delta d_c = 1.5 \times \langle \Delta d_i \rangle$ (where $\langle \Delta d_i \rangle$ is the average separation in a sample) as the threshold separation (marked with dashed lines in Fig. \ref{fig:numberDensityEvolution}A) marking the boundaries between clusters.
Using $\Delta d_c$ we estimate the evolution of the number of clusters in dumbbell suspensions at different $\Phi$ (Fig. \ref{fig:numberDensityEvolution}B). 
We observe that for small $\Phi$, all the dumbbells are uniformly distributed and they remain so even at large times. 
However, for $0.5 \le \Phi \le 0.8$, even for starting with homogeneous distributions of the dumbbells (top panel in Fig. \ref{fig:numberDensityEvolution}B), they rearrange themselves into several clusters (middle and bottom panels in Fig. \ref{fig:numberDensityEvolution}B).
For $\Phi \ge 0.9$, however, all the dumbbells appear to move as a part of a single cluster. 
This can be attributed to the small availability of the free space between adjacent dumbbells. 
As shown earlier (section \ref{sec:pair_of_active}), in a pair, two dumbbells move relative to each other if the oscillatory phase difference satisfies $\sin \theta \ne 0$. 
Therefore, an occurrence of clusters of size $N_c>n$ requires at least $n$ consecutive dumbbell pairs to have $\sin \theta_{i,i+1}<0$, where $\theta_{i,i+1}$ is the oscillatory phase difference between $i$th and $(i+1)$th dumbbells. 
Starting with randomized initial conditions, an observation of $n>7$, as is the case in Fig. \ref{fig:numberDensityEvolution}B, points towards interactions of oscillatory phases of the adjacent dumbbells. 
Therefore, we analyzed the oscillatory phases and their possible interactions in the dumbbells at different densities. 
\subsection{Phase locking in collective dynamics}
% As described in section \ref{sec:singleDumbbell}, individual dumbbell demonstrates limit cycle oscillations in length and bound myosin concentration (Fig. \ref{fig:singleDumbbellDynamics}).  

\begin{figure}[H]
    \centering
    \includegraphics[width=0.99\textwidth]{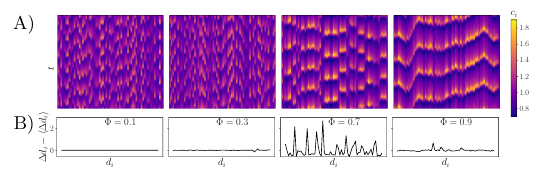}
    \caption{{\bf Density-dependent phase locking. } (A) Representative kymographs showing the concentration of bound myosin in the dumbbells for suspension densities $\Phi=0.1$, $0.3$, $0.7$, and $0.9$. 
    With an increase in suspension density, myosin dynamics in the dumbbells becomes coordinated demonstrating phase-locking at high values of $\Phi$.
    (B) Relative separation between adjacent dumbbells for the samples shown in (A). 
    The large number of peaks for $\Phi=0.7$ indicates a large number of clusters in that sample. 
    }
    \label{fig:bound_myosin}
\end{figure}

The hydrodynamic interactions among the dumbbells are expected to influence their limit cycle oscillatory dynamics (see equation \eqref{eq:dradt}). 
We looked at the concentration of bound myosin in each dumbbell for suspensions of different densities. 
The kymographs of myosin concentrations are plotted in figure \ref{fig:bound_myosin}A. 
We observe that for dilute dumbbell concentrations %($\Phi < 0.5$) 
the oscillations in the myosin concentrations are uncorrelated and the dumbbells oscillate at their respective limit cycle frequency. 
However, at higher dumbbell concentrations%($\Phi>0.5$)
, the changes in the myosin concentrations are coordinated. 
This coordinated change leads to phase-locking of the dumbbell dynamics in the suspension where all the dumbbells oscillate at a single frequency. 
During this phase-locked regime, even though all the dumbbells oscillate at the same frequency there still are some phase differences among them (last two panels in Fig. \ref{fig:bound_myosin}). 
The hydrodynamic interactions with this phase difference (see equation \eqref{eq:pairVsum}) give rise to the collective motility of the dumbbells at high concentrations. 

Further, we also observe a correspondence between phase locking and clusters in the suspensions. 
For example, we see that the dumbbells belonging to a single cluster also have their oscillatory phases locked (Fig. \ref{fig:bound_myosin}B). 
At low concentrations, there is an absence of both cluster formation as well as coordination in the dumbbell oscillations ($\Phi=0.1$, $0.3$).
For $\Phi=0.9$, we observe that all the dumbbells in the suspension oscillate in a coordinated manner (visibly continuous bands in Fig. \ref{fig:bound_myosin}A) and it reflects as a single cluster in Fig. \ref{fig:bound_myosin}B. 
On the other hand, for $\Phi=0.7$ we observe several clusters in the sample (Fig. \ref{fig:bound_myosin}B) which are also visible in the corresponding kymograph in the form of discontinuous bands. 

Therefore, we can see that at lower suspension densities there is no coordinated activity in the dumbbells but it starts to appear in the form of discrete clusters as the concentration is increased. 
As the concentration is increased further, it eventually results in a single cluster spanning the complete length of the one-dimensional suspension. 
Fig. \ref{fig:bound_myosin} shows only a few representative steady states of the dynamics of the dumbbell suspensions. 
In principle, for each $\Phi$ value there is a possibility of multiple stable steady states. 
\subsubsection{Mechanism of phase locking}
In this section, we explore the nature of steady-state oscillatory dynamics of the dumbbell suspension and different modes of phase locking due to hydrodynamic interactions.
For this, instead of simulating suspensions with a large number of dumbbells, we focused on a pair of active dumbbells. 
The two dumbbells in the pair are placed at a distance corresponding to different values of $\Phi$. 
During their numerical simulations, the velocity of the dumbbell center of mass is subtracted from the velocities of the beads of the respective dumbbells. 
This ensures that the relative positions of the dumbbells do not change over time and the effective $\Phi$ remains fixed. 
With this setup, we explored the nature of oscillatory dynamics of the pair of dumbbells at different $\Phi$. 
To quantitatively study the effect of hydrodynamic interactions on dumbbell dynamics, we looked at correlation \citep{ross2009probability} and mutual information \citep{polyanskiy2014lecture} between the oscillatory phases of the two dumbbells.

To calculate these quantities, first, we obtained marginal and joint probability density functions (PDF), $P_i(c_i)$ ($i \in \{1,2\}$) and $P_{12}(c_1,c_2)$, respectively, from the time evolution of bound myosin concentrations.
For marginal PDF of $c_i$ we divided the range of bound myosin levels into small bins and calculated the frequency of a particular value over a large duration of the steady state of numerical simulation. 
For joint PDF, we performed the same procedure as that for marginal PDF but for each pair of possible values of $c_1$ and $c_2$. 
% Details of the process are described in the Appendix. 

\begin{figure}[H]
    \centering
    \includegraphics[width=1\textwidth]{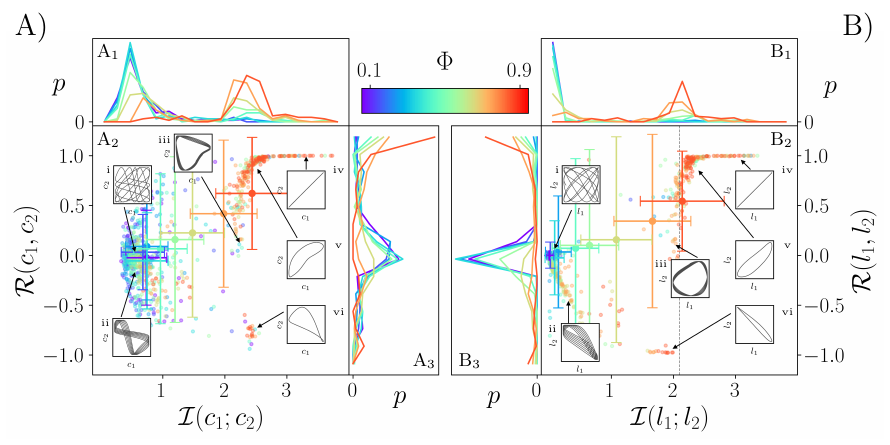}
    \caption{
    {\bf Modes of phase locking and its dependence on suspension density $\Phi$.}
    (A) Mutual Information $\mathcal{I}(c_1;c_2)$  and correlation $\mathcal{R}(c_1,c_2)$ between bound myosin concentrations in two dumbbells at fixed separations. 
    In (A$_2$) the circular markers represent different realizations of the dumbbell pair with randomized initial conditions. 
    The markers with the error bars represent average values of $\mathcal{I}$ and $\mathcal{R}$ for different $\Phi$.
    Probability distributions of $\mathcal{I}(c_1; c_2)$ and $\mathcal{R}(c_1,c_2)$ for different $\Phi$ values are in A$_1$ and A$_3$ panels, respectively.  
    (B) Mutual Information $\mathcal{I}(l_1;l_2)$  and correlation $\mathcal{R}(l_1,l_2)$ between lengths of the two dumbbells. 
    (B$_1$) and (B$_3$) show the probability distributions of $\mathcal{I}(l_1; l_2)$ and $\mathcal{R}(l_1,l_2)$ for different $\Phi$ values, respectively. 
    Insets in A$_2$ and B$_2$ show representative trajectories in $(c_1, c_2)$ and $(l_1,l_2)$ spaces, respectively. 
    In all panels, colors represent different $\Phi$ values as shown in the colorbar. 
    }
    \label{fig:r_vs_MI}
\end{figure}

Using these PDFs, we define 
mutual information between bound myosin levels in two dumbbells as 
\begin{align}
\mathcal{I}(c_1; c_2) = \int \limits_{c_1}  \int \limits_{c_2} P_{12}(c_1, c_2) \log \left[ \frac{P_{12}(c_1, c_2)}{P_1(c_1) P_2(c_2)} \right] dc_1 dc_2.
\end{align}
% where $P_{12}(c_1, c_2)$ is the joint probability distribution of $c_1$ and $c_2$ and $P_i(c_i)$  is the marginal probability distribution of $c_i$. 
Since mutual information is a measure of the mutual dependence between two variables, we expect it to be high for dumbbell pairs oscillating at the same angular frequency, also known as phase locking (for an example, see panels (iv) in Figs. \ref{fig:r_vs_MI}A$_2$ and B$_2$).  
In the absence of any phase locking between the two dumbbells, we expect $\mathcal{I}(c_1; c_2)$ to be very small (for an example, see panels (i)  Fig. \ref{fig:r_vs_MI}A$_2$  and B$_2$).

Similarly, we define the correlation between the bound myosin concentrations of two dumbbells as 
\begin{equation}
    \mathcal{R} (c_1, c_2) = \frac{\text{cov}(c_1,c_2)}{\sigma_1 \sigma_2}
\end{equation}
where 
\begin{equation}
    \text{cov}(c_1, c_2) = \int_{c_1} \int_{c_2} \left( c_1 - \mu_1\right)\left( c_2 - \mu_2\right) P_{12}(c_1,c_2) dc_1 dc_2
\end{equation}
is the covariance between $c_1$ and $c_2$, and 
\begin{align}
    \mu_i &= \int_{c_i} c_i P_i(c_i) dc_i \\
    \sigma_i^2 &= \int_{c_i} \left(c_i-\mu_i\right)^2 P_i(c_i) dc_i
\end{align}
for $i=1$,$2$ are the mean and variance of the bound myosin concentrations in two dumbbells, respectively. 
With this definition, we expect to observe $\mathcal{R} = 1$ and $\mathcal{R} = -1$ for pairs of dumbbells with synchronized oscillatory dynamics with phase differences $\theta=0$ and $\pi$, respectively (for an example, see panels (iv) and (vi) in both Figs. \ref{fig:r_vs_MI}A$_2$  and B$_2$). 
For a pair with phase-locked dynamics but with finite phase difference, $|\mathcal{R}| < 1$ and for no phase locking $\mathcal{R} \sim 0$ (for an example, see panels (i) in Fig. \ref{fig:r_vs_MI}A$_2$  and B$_2$). 

Please note that the use of any one of these two measures - correlation and mutual information - is not sufficient to completely describe the coordination between the dynamics of the two dumbbells. 
For example, if we use only mutual information, we will not be able to differentiate among the cases of phase-locked dynamics with finite phase differences (both with high values of $\mathcal{I}$, compare panels (v) and (vi) in both Figs. \ref{fig:r_vs_MI}A$_2$ and \ref{fig:r_vs_MI}B$_2$).  
Similarly, the use of only correlation will classify uncorrelated and phase-locked dynamics with $\pi/2$ phase difference to be the same (both with $\mathcal{R} \sim 0$, compare panels (i) and (iii) in both Figs. \ref{fig:r_vs_MI}A$_2$ and \ref{fig:r_vs_MI}B$_2$). 
Therefore, we used both of these quantities for the complete characterization of dumbbell interactions. 

Fig. \ref{fig:r_vs_MI}A shows mutual information $\mathcal{I}(c_1; c_2)$ and correlation $\mathcal{R}(c_1,c_2)$ between two dumbbells placed at different separations from repeated numerical simulations. 
For $\Phi \le 0.4$, we observe that the mutual information is minimal and correlation is also close to zero (see dense scatter plots on the left side of Fig. \ref{fig:r_vs_MI}A$_2$). 
Both of these quantities also show unimodal distributions with peaks at $\mathcal{I}<1$ and $\mathcal{R} \sim 0$ (Figs. \ref{fig:r_vs_MI}A$_1$ and \ref{fig:r_vs_MI}A$_3$) for $\Phi<0.4$. 
With an increase in $\Phi$ beyond $0.4$, we observe a gradual increase in the mutual information indicating phase locking. 
Similarly, an increase in $|\mathcal{R}|$ shows a decrease in the phase difference with increasing $\Phi$.
For moderate values of $\Phi$, the distributions of both $\mathcal{I}$ and $\mathcal{R}$ show a bimodal nature in Figs. \ref{fig:r_vs_MI}A$_1$ and \ref{fig:r_vs_MI}A$_3$. 
This implies that for moderate values of $\Phi$ there are two steady states of the system - one where the two dumbbells are phase-locked and another with no coordination between their dynamics. 
For high suspension densities, $\Phi=0.8$ and $0.9$, we observe a high value of $\mathcal{I}$ and $\left|\mathcal{R}\right| \approx 1$ which demonstrate that at that concentration the two dumbbells are in perfect synchronization with each other. 
Expectedly, the distribution of $\mathcal{I}$ values for large suspension densities shows unimodal distribution with a peak at $\mathcal{I} \approx 2.5$. 
However, for the correlation we again see bimodal distributions with peaks at $\mathcal{R} \approx \pm 1$. 
This shows that at high concentrations the dumbbells are always phase-locked but the exact nature of locking - in-phase synchronization or anti-phase - depends on the systems initial condition.
% The probability distribution plots in Fig. \ref{fig:r_vs_MI}A$_1$ and A$_3$ show the nature of this transition with increasing $\Phi$. 
% We can see unimodal distributions for $\mathcal{R}$ (centered around $0$) and $\mathcal{I}$ (centered at $<1$) for small values of $\Phi$. 
% Similarly, at high suspension densities, we observe unimodal distributions but peaked at $|\mathcal{R}| \sim 1$ and $\mathcal{I} \sim 3$. 
% On the other hand, for moderate values of $\Phi$ (in particular $\Phi=0.5$ in Fig. \ref{fig:r_vs_MI}A$_1$ and A$_3$), we see bimodal distributions of $\mathcal{R}$ and $\mathcal{I}$. 
% This indicates that at moderate suspension density, depending on the initial conditions, some of the suspension instances can show phase-locking but not necessarily all. 

We also calculated mutual information and correlation between the lengths of the two dumbbells and the results are shown in Fig. \ref{fig:r_vs_MI}B. 
The dumbbell lengths also show a transition from asynchronous dynamics to a phase-locked one as the suspension density is increased. 
For dumbbell lengths, we also observe a consistent decrease in the correlation $\mathcal{R}$ for $0.4 \le \Phi \le 0.7$ indicating phase-locking with opposing phases. 
That is, for this range of $\Phi$ when one dumbbell is stretching the other one is contracting, albeit not exactly in the opposite phase. 
This negative correlation between dumbbell lengths persists till $\Phi = 0.7$.
As $\Phi$ is increased further, we can see that the correlation jumps from $\mathcal{R} = -1$ to $\mathcal{R}=1$ (marked with a dashed line in Fig. \ref{fig:r_vs_MI}B$_2$ with a continuous increase in $\mathcal{I}$.
At very high $\Phi$ values the dynamics of the dumbbell lengths shows $\mathcal{R}=1$ and high values of $\mathcal{I}$ indicating perfect synchronization between the lengths of the two dumbbells. 
Similar to the bound myosin levels, the correlation and mutual information between the lengths of the two dumbbells also transition from unimodal distributions to bimodal to unimodal again (Fig. \ref{fig:r_vs_MI}B$_1$ and B$_3$).

\section{Discussion}
In this work, we have demonstrated three properties of the suspensions of active but sessile dumbbells- first, the transition from sessile to collective motion, second, hydrodynamics-induced phase separation, and third, phase-locking of the dumbbell activity. 
All of these phenomena are dependent on the suspension density and driven by hydrodynamic interactions among the dumbbells. 
It is not surprising to observe the emergence of the collective motion of a pair of dumbells (if phases are different) as a dumbbell-pair is not very different from the bead-spring models of microwimmers \citep{C4SM02611J,  PhysRevE.97.023102, rizvi2018size}. 
In fact, in the absence of the interactions between the myosin activity across all dumbbells, the time periods of each dumbbell take a fixed value and the system can be analyzed analytically (albeit the calculations can be tedious).
In section \ref{sec:pair_of_active} we have taken advantage of this fact and obtained analytical expressions for the velocity of the dumbbells. 
The numerical simulations, however, suggest that ignoring interactions between the myosin activity results in incorrect conclusions about the density dependence on collective velocity (Fig. \ref{fig:slopeandvelocity}B). 
We observe that hydrodynamic interactions between dumbbells result in their relative and collective motion, and the relative motion influences the coupling of their cytoskeletal dynamics. 
Therefore, the emergence of collective motility of the suspension of sessile dumbbells is an outcome of a two-way coupling among them.

For low suspension density, the collective displacement of the dumbbells is negligible. 
Even though in this work we have ignored the effect of diffusion, we expect that at low dumbbell density, the collective transport will be primarily diffusion-driven. 
As we see here, at high suspension densities the collective motion is observed and for the parameter values explored in this work, the velocity of the collective motion is found to be as high as of the order of $10^{-2}$ (non-dimensional units, see Fig. \ref{fig:slopeandvelocity}B). 
For a fibroblast cell of size $20$ $\mu$m and oscillation time period of $10$ s, this non-dimensional collective velocity translates to $10^{-1}$ $\mu$m/min. 
This magnitude of the velocity is comparable to the slow mesenchymal mode of cell motility \citep{pavnkova2010molecular}. 
It has to be noted that in mesenchymal cell motility, the cells are polarized and take help from the surrounding medium by generating traction forces with the help of adhesive interactions. 
In contrast, the model of the cell presented in this work lacks polarity and does not have any adhesive interaction with the medium. 

Although the density-dependent transition to collective movement in the system studied here resembles that observed in flocking models, the underlying mechanisms driving the transition in the two systems are fundamentally different.
In the flocking models, the emergence of collective movement of the active agents is due to their alignment interactions \citep{ramaswamy2010mechanics} that are ingrained in the model itself.
In the dumbbell models, albeit studied in 1D here, the interactions are hydrodynamic and their consequences manifest in the form of phase-locking of the cytoskeletal dynamics, and relative and collective movements. 
Similarly, the collective motion reported here is also different from that for the sessile \emph{Caulobacter crescentus} which actively disperse through the flagellar motors of motile members \citep{zeng2020self}. 
Due to the presence of an actively motile fraction of the population, this system shows a relatively higher collective velocity of the order of $1$ $\mu$m/min as opposed to $0.1$ $\mu$m/min for a purely sessile population studied in this work.

We also observe hydrodynamics-induced phase separation for sessile dumbbell suspension which results in regions of high and low densities of dumbbells in the system. 
The coordination within a cluster of dumbbells (characterized in terms of phase similarity) is higher as compared to two dumbbells from different clusters. 
This results in a high collective translational velocity of a cluster. 
Since the emergent polarity (direction of movement) of each cluster is independent the average translational velocity of the system as a whole takes a lower value. 
Phase separation has also been observed in other active suspensions, such as systems of active Brownian particles. 
However, due to the motile nature of the individual agents in these systems, the phase separation arises from a different mechanism, known as motility-induced phase separation (MIPS) \citep{cates2015motility}.
In MIPS, when self-propelling particles accumulate in high-density regions, their motility decreases, leading to a feedback loop that drives phase separation into dense and dilute regions, even in the absence of attractive interactions.
On the other hand, in sessile dumbbells, the phase separation is due to hydrodynamic attraction/repulsion between the dumbbells which depends on their relative phases. 

It needs to be noted here that the results shown in this paper
are for a one-dimensional model. In higher dimensions (two or three) we expect that the qualitative features of the system - density-dependent transition from sessile to collectively motile states - will not only hold but will show even richer behavior. 
In two and three dimensions we can borrow terminology from the nematic liquids to describe the dynamics of the system. 
Another limitation of the current model is the simplistic representation of the cell by a bead-spring dumbbell. 
In more realistic representations, one can model a biological cell with an elastic capsule with activity leading to oscillatory shape changes \citep{farutin2013amoeboid}. 
Even though here we have looked at the effect of hydrodynamic interactions on collective dynamics for sessile cells, the model can also be easily extended to motile cells. 
A straightforward generalization can be to use three-bead microswimmers of linear \citep{C4SM02611J} and triangular \citep{PhysRevE.97.023102} shapes. 
We hope to report on some of these aspects in the future. 

\section{Conclusions}
In this study, we have explored the collective dynamics of suspensions of active but sessile dumbbells and demonstrated three key phenomena: the density-dependent transition from sessility to collective motion, hydrodynamics-induced phase separation, and phase-locking of dumbbell activity. 
These phenomena underscore the critical role of hydrodynamic interactions, which drive the relative and collective movements of the dumbbells while coupling their cytoskeletal dynamics. 
% Unlike traditional models of collective motion driven by alignment interactions, such as flocking systems, the mechanisms here are rooted in hydrodynamics and exhibit unique features, including emergent phase-locking and density-dependent motility. 
For low suspension densities, motion is minimal and likely diffusion-driven, while at higher densities, collective velocity becomes significant, reaching magnitudes comparable to the slow mesenchymal mode of cell motility. 
Notably, this collective motion is achieved without polarity or adhesive interactions with the medium.

In summary, our work highlights the crucial role of hydrodynamic interactions on the collective behavior of active suspensions, providing a basis for understanding similar phenomena in biological and synthetic systems. 
We hope these insights will inspire further studies and applications in understanding collective cell motility in more biologically realistic settings, and the design of microfluidic devices.

\section*{Acknowledgements}
The financial support for this work was provided by a seed grant from the Indian Institute of Technology Hyderabad and a Startup Research Grant (SRG) from SERB, India. 

% \section{Appendix}
% \begin{align}
%     \dot{c}_i &= - \frac{1}{\tau_i} \left( c_i - c_{0i}\right) - \frac{c_i}{l_i} \dot{l}_i \\
%     f_i &= t_1 (c_i - c_{0i}) + k_1 (l_i - l_0) + k_3 (l_i - l_0)^3 \\
%     \dot{l}_1 &= 2 \left( \mu - \frac{2G}{l_1}\right)f_1 + 4G f_2 \left[ \frac{1}{2d + l_1 + l_2} - \frac{1}{2d + l_1 - l_2} + \frac{1}{2d - l_1 - l_2} - \frac{1}{2d - l_1 + l_2}\right] \\
%     \dot{l}_2 &= 2 \left( \mu - \frac{2G}{l_2}\right)f_2 + 4G f_1 \left[ \frac{1}{2d + l_1 + l_2} - \frac{1}{2d + l_1 - l_2} + \frac{1}{2d - l_1 - l_2} - \frac{1}{2d - l_1 + l_2}\right]
% \end{align}

% \bibliographystyle{unsrtnat}
\bibliographystyle{plainnat}
\bibliography{reference}
\end{document}